\documentclass[lettersize,journal]{IEEEtran}
\usepackage[utf8]{inputenc}

\usepackage{verbatim}
\usepackage{color}
\usepackage{multicol}
\usepackage{placeins}
\usepackage{float}
\usepackage{graphicx}
\usepackage{cite}

\usepackage[caption=false,font=footnotesize]{subfig}

\usepackage{xcolor}

\usepackage{graphicx}

\usepackage{algorithmic}
\usepackage{algorithm}

\usepackage{booktabs}
\usepackage{enumitem}

\usepackage{amsmath}
\usepackage{amssymb}
\usepackage{amsthm}
\usepackage{bm}

\newcommand{\transpose}{^{\mkern-1.5mu\mathsf{T}}}
\newcommand{\hermitian}{^{\mathsf{H}}} %Hermitian transpose
\DeclareMathOperator*{\argmin}{argmin}

\newcommand{\norm}[1]{\left\lVert #1 \right\rVert}
\newcommand{\set}[1]{\left\{#1\right\}}
\newcommand{\reals}{\mathbb{R}}
\newcommand{\complexs}{\mathbb{C}}

\newcommand{\pilot}{\bm{\phi}}
\newcommand{\channel}{\bm{g}}
\newcommand{\noise}{\bm{N}}
\newcommand{\Y}{\bm{Y}}
\newcommand{\y}{\bm{y}}
\newcommand{\x}{\bm{x}}
\newcommand{\A}{\bm{A}}
\newcommand{\taup}{\tau_{\text{p}}}
\newcommand{\balpha}{\bm{\alpha}}

\newcommand{\abs}[1]{\left|#1\right|}
\newcommand{\CN}[2]{\mathcal{CN}\left( #1,#2 \right)}

\newcommand{\sumusers}{\sum_{k=1}^{K}}

\title{Activity Detection in Distributed Massive MIMO With Pilot-Hopping and Activity Correlation}
\author{\IEEEauthorblockN{Ema~Becirovic,  Emil~Bj\"{o}rnson and Erik~G.~Larsson} 
	\thanks{This work was supported in part by ELLIIT and in part by KAW foundation.} 
	
	\thanks{E. Becirovic and E.~G.~Larsson are with the  Department of Electrical Engineering (ISY), Link\"{o}ping University,  Link\"{o}ping, Sweden (e-mail: \{ema.becirovic, erik.g.larsson\}@liu.se). E.~Bj\"ornson is with the Dept. of Computer Science, KTH Royal Institute of Technology, Kista, Sweden (e-mail: emilbjo@kth.se). }%
}

\begin{document}

	\maketitle

	\begin{abstract}
		
		Many real-world scenarios for massive machine-type communication involve sensors monitoring a physical phenomenon. As a consequence, the activity pattern of these sensors will be correlated. In this letter, we study how the correlation of user activities can be exploited to improve detection performance in grant-free random access systems where the users transmit pilot-hopping sequences and the detection is performed based on the received energy. We show that we can expect considerable performance gains by adding regularizers, which take the activity correlation into account, to the non-negative least squares, which has been shown to work well for independent user activity. 
				
	\end{abstract}

	\begin{IEEEkeywords}
		Distributed massive MIMO, grant-free random access, correlated activity detection
	\end{IEEEkeywords}

	\section{Introduction} 
	
	\IEEEPARstart{M}{assive} machine-type communications (mMTC) is one of the core use cases of 5G \cite{ituvision2015}. 
	mMTC refers to scenarios where many devices are sending intermittent data. This creates a vast load on the random access protocols.
	Random access for mMTC has been studied in many works recently \cite{Bockelmann18,Chen21} and especially a large focus has been on grant-free random access since the communication overhead will be smaller when there is no contention resolution \cite{confver,Liu18}.
	In grant-free random access, pilots are used to both detect the active users and estimate their channels. Since, there is a huge number of devices in the system, they cannot be assigned orthogonal pilots in each coherence interval. There are two strategies to solve this problem: the first is to assign non-orthogonal pilots \cite{Liu18}, and the second is to assign orthogonal pilots in each coherence interval, but have pilot-hopping sequences that span multiple coherence intervals \cite{deCarvalho17a,confver,Sarband21}.
	
	One important relief of random access for mMTC is that, even though there is a huge number of devices, only a small fraction of them are active at a given time since they are only active when they have information to transmit. This fact makes it justifiable to cast the activity detection problem as a compressed sensing problem and solve it with algorithms that have been proven efficient in such scenarios.
	
	In addition to assuming sparsity, additional side-information can be used to further improve the detection performance. One such example is temporal correlation of the user activity: a user is probably active in many consecutive time slots since it is active until it has transmitted all the data \cite{Zhu21,WangQ22}. Another type of side-information that can be exploited is correlation of the channel between the active user and the base station. This channel can be spatially correlated \cite{Djelouat21,Djelouat22} or both spatially and temporally correlated \cite{Cheng21}. Such side-information is used to improve methods which jointly detect the active users and estimate their channels. 
		
	If the activity of the devices is governed by a physical phenomenon, we can presume that the \emph{activity} of devices monitoring the same phenomenon is correlated. 
	The previously mentioned grant-free random access algorithms \cite{Zhu21,WangQ22,Djelouat21,Djelouat22,Cheng21} and protocols exploit side-information. However, using correlated user activity as side-information has been somewhat neglected for grant-free random access.
	One work that considers correlated user activity is \cite{Stern19}, which considers an unsourced random access framework. In that work, the devices can either transmit standard messages, or alarm messages. 
	The alarm messages depend on a physical phenomenon and all devices will transmit the same message in the case of an alarm.
	Recently, \cite{Liu2022} considered correlated user activities in a grant-based random access setting. They found the preamble (pilot) selection distribution that minimizes preamble collisions given a correlated activity distribution.

\begin{figure}
	\centering
	\includegraphics{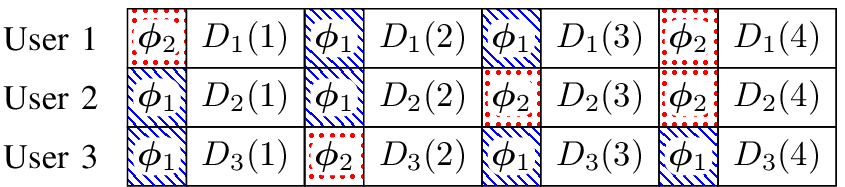}
	\caption{Three users transmitting pilot-hopping sequences of length four using two pilots, $\bm{\phi}_1$ and $\bm{\phi}_2$. Additionally, in coherence interval $t$ the user $k$ also transmits uplink data $D_k(t)$.  }
	\label{fig:coherence-interval-RA-MaMi}
\end{figure}

	In this paper, we generalize the transmission model proposed in  \cite{confver,deCarvalho17a} and implemented in \cite{Sarband21} where the devices transmit pilot-hopping sequences in order to aid in activity detection and channel estimation, see Fig.~\ref{fig:coherence-interval-RA-MaMi}. A single-cell massive multiple-input-multiple-output (MIMO) system where the user activities are independent was studied in \cite{confver}, while in this paper, we consider a distributed massive MIMO system with correlated user activities. The technical challenge is to formalize a detection problem that can adequately capture the activity correlation. We address this issue by solving a regularized non-negative least squares (NNLS) problem, where the regularizer takes the activity correlation into account. To the best our knowledge, this is the first work to consider correlated user activities in grant-free random access.

\section{Distributed Massive MIMO System Model}
We consider a distributed MIMO system where $L$ base stations are equipped with $M$ antennas each and collectively serve $K$ single-antenna users. Note that when $L=1$, this is the standard co-located single-cell massive MIMO system model. Not all of these users are active. The activity is modeled as random but there might be correlation between users. We assume a block fading model; the channel is assumed to be time invariant and frequency flat in each coherence interval. 

We study a grant-free access transmission model where each communication round spans $T$ coherence intervals. In each coherence interval there is a pilot phase, which is $\taup$ symbols long, such that there are $\taup$ orthogonal pilots, and a data phase, which spans the rest of the coherence interval. The users are assigned a unique \emph{pilot-hopping sequence} wherein each of the $T$ coherence intervals one of the $\taup$ orthogonal pilots is chosen, see Fig.~\ref{fig:coherence-interval-RA-MaMi}.
While the pilots might collide in individual coherence intervals, the pilot collisions get averaged over the whole pilot-hopping sequence and data can still be reliably transmitted \cite{deCarvalho17a}.
Hence, with highly frequency-selective channels, where $\taup$ will be small due to a limited coherence bandwidth, more collisions will occur per coherence~block. 

At the pilot phase 
of coherence interval $t$, the base stations \emph{collectively} receive the $ML\times\taup$ signal
\begin{align}
\Y^t &= \sum_{k = 1}^K\sum_{j = 1}^{\taup}\alpha_kS_{j,k}^t\sqrt{\taup p_k}\channel_{k}^t\pilot_j\hermitian + \noise^t, \; t = 1, \dots, T
\label{eq:receivedsignal} \\
&\hspace{-16pt}\text{where} \qquad\qquad \alpha_k = \begin{cases}1,& \text{if user }k\text{ is active,}\\ 0, &\text{otherwise,} \end{cases}  \\
S_{j,k}^t &= \begin{cases}
1 & \text{ if user }k\text{ sends pilot }j\text{ at pilot phase }t\text{,} \\
0 & \text{ otherwise,}
\end{cases} 
\end{align}
$p_k$ is the transmit power of user $k$, $\channel_{k}^t \in \complexs^{ML}$ is the collective channel between user $k$ and the base stations in coherence interval $t$, and $\pilot_j \in \complexs^{\taup} $ is the $j$:th pilot consisting of $\taup$ symbols. The pilots are mutually orthogonal, $\pilot_i\hermitian\pilot_j = 0, i\neq j$, and have unit norm, $\norm{\pilot_i} = 1$. Finally, $\noise^t \in \complexs^{ML \times \taup}$ is noise with i.i.d. $\CN{0}{\sigma^2}$ elements.
The received signal in \eqref{eq:receivedsignal} can be viewed as blocks of received signals from each base station,
\begin{equation}
\Y^t = 
\begin{bmatrix}
(\Y^t_1)\transpose&\dots&(\Y^t_l)\transpose&\dots&(\Y^t_L)\transpose
\end{bmatrix}\transpose,
\end{equation}
where 
$\Y_l^t \in \complexs^{M\times\taup}$ 
is the received signal at base station $l$ at the pilot phase of coherence interval $t$.
The model assumes that the users are synchronized in the sense that all the active users start their pilot-hopping sequence at the same coherence interval. Hence, the same users are active during the $T$ coherence intervals in \eqref{eq:receivedsignal}. If new users want to access the network, they would need to wait until the next pilot-hopping sequence starts.

\subsection{Asymptotic Energy System Model}
In each coherence interval, an estimate of the received signal energy over each pilot is computed as 
\begin{equation}
E_{i,t} = \frac{(\Y^t\pilot_i)\hermitian(\Y^t\pilot_i)}{ML}-\sigma^2 = \frac{\norm{\Y^t\pilot_i}^2}{ML}-\sigma^2,
\end{equation}
which is a sufficient statistic for the user activity.
These energy estimates are used to detect the active users.
Assuming that the channels have the asymptotic channel hardening property, i.e., 
\begin{equation}
\frac{\norm{\channel_k^t}^2}{ML}\to\beta_k,\quad\text{ as }\quad ML\to\infty,\label{eq:channel-hardening}
\end{equation} where $\beta_k$ is the large-scale fading coefficient\footnote{In distributed systems, the large-scale fading coefficient should be interpreted as the mean of the large-scale fading coefficient to all base stations, which usually depend on the distance to each base station, i.e., $\beta_k = \frac{1}{L}\sum_{l=1}^L\beta_k^l$, where $\beta_k^l$ is the large-scale fading coefficient between user $k$ and base station $l$. In this case, the channel hardening property also depends on the distribution of $\beta_k^l$. Here it is implied that $\beta_k$ is the same for all $t$ but this assumption is not necessary as long as $\beta_k$ is known.}, and asymptotic favorable propagation property, i.e., 
\begin{equation}
\frac{(\channel_k^t)\hermitian(\channel_{k'}^t)}{ML}\to0,\quad\text{ as }\quad ML\to\infty, k\neq {k'}, \label{eq:favorable-propagation}
\end{equation} 
we see that	as $ML\to \infty$, 
\begin{equation}
E_{i,t} = \frac{\norm{\Y^t\pilot_i}^2}{ML}-\sigma^2 \rightarrow \sum_{k = 1}^K\alpha_kS_{i,k}^t\taup p_k\beta_k.
\label{eq:energylimit}
\end{equation}
Channel hardening and favorable propagation are common in many propagation scenarios in massive MIMO and are fulfilled for e.g. Rayleigh fading \cite[Ch.~7]{redbook} \cite[Ch.~2]{massivemimobook} and uniform random line-of-sight channels \cite[Ch.~7]{redbook}.
In practice, the total number of antennas $ML$ will be finite and \eqref{eq:energylimit} can then be interpreted as an approximation. For the co-located case ($L~=~1$) with i.i.d. Rayleigh fading, the approximations are normally tight at around $M=50$ antennas since the variance of \eqref{eq:channel-hardening} and \eqref{eq:favorable-propagation} scale with $\frac{1}{M}$ \cite[Sec.~2.5]{massivemimobook}.

We introduce the vector notation 
\begin{gather}
\balpha = \left[ \alpha_1,\dots,\alpha_K\right]\transpose, \\
%\end{gather}
%\begin{gather}
\hspace{-9pt}\text{and }\y = \left[ E_{1,1}, \dots, E_{\taup,1},\dots,E_{i,t},\dots E_{1,T},\dots,E_{\taup,T} \right]\transpose.
\end{gather} 
Further, we introduce the $\taup T~\times~K$ matrix, 
\begin{equation}
\setlength{\arraycolsep}{0pt}
\renewcommand\arraystretch{0}
\A = 
\begin{pmatrix}
S_{1,1}^1\taup p_1\beta_1 &&\cdots &&S_{1,K}^1\taup p_K\beta_K \\
\vdots&&\ddots&&\vdots\\
S_{\taup ,1}^1\taup p_1\beta_1&&\cdots&&S_{\taup ,K}^1\taup p_K\beta_K \\
&\ddots&&&\\
\vdots&&S_{i,k}^t\taup p_k\beta_k&&\vdots \\
&&&\ddots&\\
S_{1,1}^T\taup p_1\beta_1&&\cdots&&S_{1,K}^T\taup p_K\beta_K\\
\vdots&&\ddots&&\vdots\\
S_{\taup ,1}^T\taup p_1\beta_1 && \cdots &&S_{\taup ,K}^T\taup p_K\beta_K
\end{pmatrix}.
\label{eq:mata}
\end{equation}	
Note that with $\taup$ pilots and $T$ coherence intervals, there are $(\taup)^T$ unique pilot-hopping sequences. Therefore, in order for each user to have unique sequences we require $K\leq (\taup)^T$. However, since we assume that there is a \emph{massive} number of users, the product of the number of pilots and the sequence length is most likely smaller than the number of users: $\taup T~\leq~K$. Hence, the matrix $\A$ is wide.
Now, we can express the limit in \eqref{eq:energylimit} using this notation:
\begin{equation}
	\y \to \A\balpha \quad \text{ as }\quad ML\to \infty. \label{eq:linearenergylimit}
\end{equation}
With a finite number of antennas, \eqref{eq:linearenergylimit} will only hold approximately. We consider the asymptotic system model to be 
\begin{equation}
	\y = \A\balpha + \bm{n}, \label{eq:asymptotic-system-model}
\end{equation}
where $\bm{n}$ is the zero mean noise that is generally not Gaussian. 

\section{User Activity Detection}
	We aim to detect the active users using the the received energies based on the model in \eqref{eq:asymptotic-system-model}. 

\subsection{Independent User Activity}\label{sec:independent-user-activity}
	First, we will treat the case where the user activities are independent. 
	Although there are fewer measurements ($\taup T$) than the number of users ($K$), we know that only a few of them are active. Hence, it is pertinent to view the active user detection problem as a compressed sensing problem.
	Minimizing the noise given that the $\alpha_k$:s are 0/1-variables gives a combinatorial problem,
	\begin{equation}
		\argmin_{\balpha \in \set{0,1}^K} \norm{\A\balpha-\y}^2,
	\end{equation}
	which is computationally costly to solve through an exhaustive search. Therefore, the problem is relaxed and we allow $\balpha$ to take any non-negative value. The relaxed problem becomes the NNLS %non-negative least-squares (NNLS)
	\begin{equation}
		\argmin_{\balpha\geq\bm{0}} \norm{\A\balpha-\y}^2. \label{eq:NNLS}
	\end{equation}
	After solving \eqref{eq:NNLS} using any convex solver, each element is thresholded to detect the active users.
	If  $\A$ in \eqref{eq:mata} is properly normalized,  it has the so-called self-regularizing property \cite[Condition~1]{Slawski2013}.
	 To show this, first note that each column of $\A$ contains exactly $T$ non-zero entries, corresponding to the pilot-hopping sequence for each user. 
	 Suppose we normalize $\A$  such that the norm of all its columns are equal.
This can be achieved by forcing the received signal-to-noise ratios  (SNR) for each user to be equal, i.e., the transmit power of each user should be inversely proportional to the large-scale fading coefficient, i.e., statistical
channel inversion \cite{massivemimobook}: $p_k = \frac{p \beta_\text{min}}{\beta_k}$, where $p\taup$ is the maximum allowed transmit power per pilot and $\beta_\text{min}=\min_k \beta_k$. 
	 Then, take $\bm{X}=\frac{\A}{\sqrt{\taup}p\beta_\text{min}}$, $\bm{w}=\left[1,1,...,1\right]\transpose$  in the definition of \cite[Condition~1]{Slawski2013},  and verify that the condition is satisfied.	
	 
	 Generally, NNLS problems with a self-regularizing measurement matrix yield  sparse solutions, even without introducing an explicit sparsity-enforcing regularizer, if the noise distribution is sub-Gaussian  \cite[Thm.~1]{Slawski2013}; see also \cite{Bruckstein08,Kueng18, Foucart14}. 
	 In the present context, with Rayleigh fading channels, the noise $\bm{n}$ is not sub-Gaussian as it contains \emph{products} of Gaussian random variables. 
	 Therefore, \cite[Thm.~1]{Slawski2013} does not strictly apply. Yet, we take the self-regularizing property of $\A$ as an explanation for why our method works and specifically yields sparse solutions in our numerical experiments.
	 
	 The NNLS approach, based on measurements of  signal energies as in \eqref{eq:asymptotic-system-model}, and formulation of the activity detection problem as an NNLS problem, was originally proposed in \cite{confver}; therein,  however, only the single-cell case   ($L=1$) was studied.

\subsection{Correlated User Activity}\label{sec:correlated-user-activity}

In this section, we propose an algorithm for activity detection with user activity correlation.
To exploit the activity correlation in the detection, the correlation needs to be known a priori or estimated to some extent. For example, the activity correlation can be estimated based on past successful transmissions. The correlation can for example arise because the devices are monitoring the same type of event or because the devices are in the same geographical area.

Our proposed approach is to add a regularizer, $R(\balpha)$, to the NNLS in \eqref{eq:NNLS} to obtain
	\begin{equation}
		\min_{\balpha\geq\bm{0}} \norm{\A\balpha-\y}^2 + R(\balpha).
	\end{equation} We consider two different regularizers. 
	\begin{LaTeXdescription}
		\item[Group-LASSO-inspired regularizer:]
		If $\balpha$ is group sparse, i.e., only a few pre-determined groups of users are active, an approach from compressed sensing is to use the $\ell_1/\ell_2$-regularizer \cite{Rish15}
		\begin{align}
		R(\balpha) &= \lambda\sum_{j=1}^G c_j\norm{\balpha_{\mathcal{G}_j}}_2,
		\end{align}
		where $G$ is the number of groups, $\mathcal{G}_j$ is the $j$:th group, $\balpha_{\mathcal{G}_j}$ denotes the sub-vector of $\balpha$ where $\mathcal{G}_j$ contains the selected components,  $c_j$ is the weight associated with the $j$:th group, and $\lambda$ is the regularization parameter which decides how much weight should be put on the regularization. If the groups are non-overlapping, this minimization problem is the group least absolute shrinkage and selection operator (LASSO) problem (with an additional non-negativity constraint) \cite[Ch.~6]{Rish15}.
		\item[Total-variation-inspired regularizer:]
		If a user is more prone to be active if its neighbors are active, a suitable regularizer is
		\begin{align}
		R(\balpha) &= \lambda \sumusers c_k \sqrt{\sum_{j\in\mathcal{N}(k)} \abs{ \alpha_k - \alpha_j}^2}, 
		\end{align}
		where $\mathcal{N}(k)$ are the neighbors of user $k$ in some correlation sense, $\balpha_{\mathcal{N}(k)}$ denotes the sub-vector of $\balpha$ where $\mathcal{N}(k)$ contains the selected components, $c_k$ is the weight associated with the $k$:th user, and $\lambda$ is the regularization parameter. This regularizer can be interpreted as the $\ell_1/\ell_2$-regularizer on the difference between users and their neighbors. The regularizer will enforce that only a few users' activities differs from their neighbors'. This regularizer is a generalization of the isotropic two-dimensional total variation denoising used for, e.g., images, where the neighbor set contains only the directly adjacent pixels \cite{Condat18}. 
	\end{LaTeXdescription}
	The particular choices of groups and neighbor sets will depend on the application. Here, we used an approach where all users in a group are treated equally, and equally contribute to the overall objective function. Another approach is to weigh the regularization such that user pairs with higher correlation have a higher impact.

	When the activity is correlated it might not be vital exactly which users are active, but instead a detection of e.g., the monitored event is more important. Therefore, performance evaluation needs to be handled with care. In the case where the correlation comes from location based information, i.e., when users are activated by nearby events, we suggest evaluating how well different algorithms perform at detecting the position of the events. Consider $E$ events occurring at positions $\bm{e}_i \in \reals^2$, $i=1,\dots,E$. After detecting the active users (by thresholding the results from the regularized NNLS), their positions are clustered into $E$ clusters. Note that, here it is assumed that we know the number of events through a genie but nothing prevents the estimation of the number of events. After clustering, we obtain estimates of the event positions $\hat{\bm{e}}_j, j = 1,\dots,E$. Next, we find the pairings, $\mathcal{P} = \{(i,j)\}$, such that 
	\begin{equation}
	\frac{1}{E}\sum_{(i,j)\in \mathcal{P}} \norm{\bm{e}_i - \hat{\bm{e}}_j}^2 
	\end{equation}
	is minimized. This gives the squared distance between the true and detected events. 
	
\section{Numerical simulations}
In this section we study a specific simulation scenario where a set of user terminals monitor the occurrence of some event. The scenario is chosen to illustrate the potential and expected gain that is obtained by including the knowledge of user activity correlation. 
We simulate a scenario with $L=4$ base stations, each with $M=32$ antennas. The simulation scenario resembles an industrial Internet-of-things setting \cite{Xu14}.

The ``world'' is a 2-dimensional plane\footnote{Note that the absolute dimensions are unimportant since the users perform statistical channel inversion power control.}, $[0,1]\times[0,1]$. There are $K=1296$ users placed on a $36\times36$ square grid. The base stations are placed at the center of each edge, i.e., at positions $(0,0.5)$, $(1,0.5)$, $(0.5,0)$ and $(0.5,1)$. Within the plane, $E=3$ events occur uniformly at random. User $k$ at position $\x_k$ is activated by event $i$ at position $\bm{e}_i$ with probability $p_{k,i} = \exp\left(\frac{-\norm{\x_k - \bm{e}_i}^2}{2\sigma_e^2}\right)$, where $\sigma_e^2 = 0.001$ is chosen such that, on average, one event approximately activates $7.5$ users.%\footnote{\mytodo{Ema: I did not calculate the exact expected value, but ran simulations...$100 000$ simulations give $7.49679$}}.

There are $\taup=10$ orthogonal pilots and the pilot-hopping sequences span $T=10$ coherence intervals.
The users' pilot-hopping
sequences are chosen uniformly at random from all the possible $(\taup)^T$
sequences. The channels are modeled as i.i.d. Rayleigh fading; the channel between user $k$ and the $l$:th base station is $\channel_k^l \sim \CN{\bm{0}}{\beta_k^l\bm{I}}$, where $\beta_k^l$ is the large-scale fading to the $l$:th base station and $\beta_k = \frac{1}{L}\sum_{l=1}^L\beta_k^l$. The large-scale fading depends on the distance to each base station: $\beta_k^l = \gamma (d_k^l)^{-3.67}$, where $d_k^l$ is the distance between user $k$ and base station $l$, $3.76$ is the path-loss exponent, and $\gamma$ is a constant. The users perform statistical
channel inversion power control as described in Section~\ref{sec:independent-user-activity}. The SNR is defined as $\mathsf{SNR} = \frac{p\beta_\text{min}}{\sigma^2}$. In the simulation we have chosen the constant $\gamma$ such that $\mathsf{SNR} = 10$~dB. 

For the total-variation-inspired (TV) regularizer, the neighbor set of user $k$ is chosen as 
\begin{equation}
	\mathcal{N}(k) = \{ i: \norm{\x_k - \x_i} < r  \},
\end{equation}
where $r=0.05$ is chosen such that, when the $k$:th user is not on the edge of the grid, there are $9$ users in the set including the $k$:th user itself. For the group-LASSO-inspired (GLASSO) regularizer, the groups are chosen to be the $K$ different neighbor sets, i.e., $G=K$ and $\mathcal{G}_k = \mathcal{N}(k)$. Further, all users and groups have the same (unit) weight, $c_k = 1$, for both the TV and GLASSO regularizers.  All three optimization problems (NNLS, TV and GLASSO) are solved with the MOSEK \cite{mosek} solver in CVXPY \cite{cvxpy}. 

The probability of missed detection, $p_\text{m}$, and the probability of false alarm, $p_\text{fa}$, are defined as 
\begin{align}
p_\text{m} &= \frac{\text{\# undetected active users}}{\text{\# active users}},\quad\text{and}\\
p_\text{fa} &= \frac{\text{\# detected inactive users}}{\text{\# inactive users}}.
\end{align}

\begin{figure*}
	\centering 
		\subfloat[TV]{
			\includegraphics{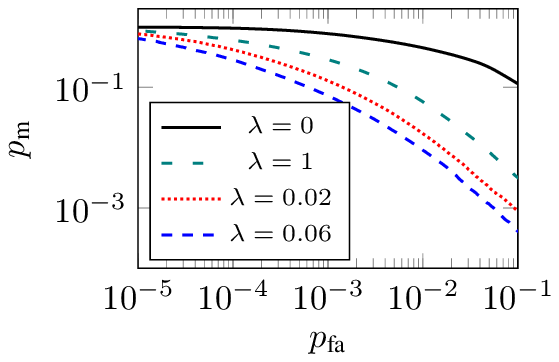}
	}
	\hfill 
	\subfloat[GLASSO]{
		\includegraphics{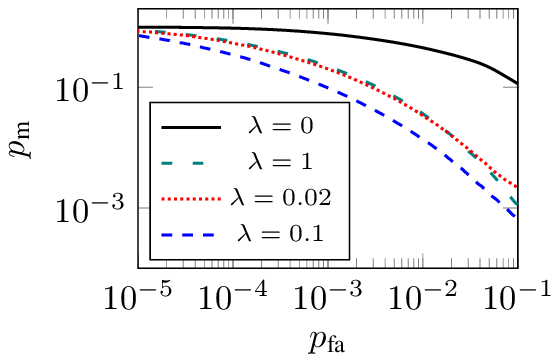}
	}
	\hfill
	\subfloat[Comparison]{
		\includegraphics{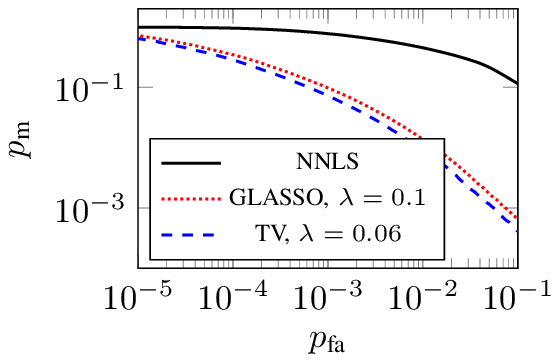}
		}
	\caption{ROC comparison of the proposed regularizers with different regularization parameters, $\lambda$. Note that, $\lambda=0$ corresponds to NNLS for both the regularizers.}
	\label{fig:ROC}
\end{figure*}

Fig.~\ref{fig:ROC} shows the receiver operating characteristic (ROC) curves for the different methods. Different false alarm and miss detection probabilities can be achieved by choosing different thresholds. We see that both the proposed methods, TV and GLASSO, perform better, i.e., at a given false alarm probability, the miss detection probability is lower, than for NNLS when the appropriate regularization parameter, $\lambda$, is chosen. We note that in this case, with these regularization parameters, TV with $\lambda=0.06$ performs best. However, the best regularizer (and regularization parameter) will depend on application and specifically how the user activities are correlated.
We conjecture that TV performs better when the activity correlation comes from the device location, as indicated by the simulations, and that GLASSO is better suited when the activity correlation is such that devices activate in predetermined groups.

\begin{figure}
	\centering
	\includegraphics{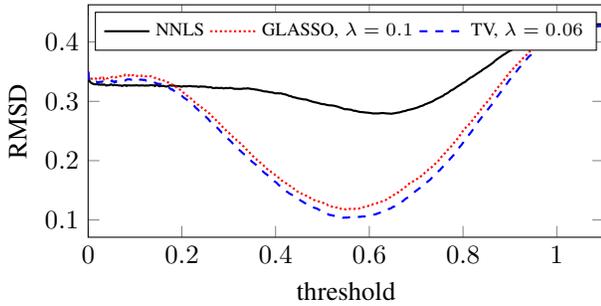}
	\caption{After the active users are detected, their positions are clustered into $E=3$ clusters and the RMSD between the cluster centers and the true event positions is reported. }
	\label{fig:root-mean-square-distance}
\end{figure}

As we have seen, the proposed detection algorithms are better at detecting active users than the baseline NNLS.
We also report the root-mean-square distance (RMSD) in Fig.~\ref{fig:root-mean-square-distance}, obtained by the process described in Section~\ref{sec:correlated-user-activity}.
After detecting the active users, their positions are clustered into $E=3$ clusters by the \mbox{K-means} algorithm \cite[Ch.~5]{datamining}. In case of no detected users, we place all events in the center of the plane.
From the figure, we see that taking the activity correlation into account gives more accurate event position estimates than NNLS, and we can deduce that the optimal (in terms of RMSD) threshold is somewhere between 0.4 and 0.7. In practice the threshold can be tested empirically and set based on application requirements. 

We have also performed simulations with fewer antennas (not pictured due to space constraints) and although the approximation in \eqref{eq:asymptotic-system-model} is worse, the conclusions follow the ones presented here; exploiting the activity correlation will give better performance.

\section{Conclusions}

We studied a user activity detection problem for mMTC users in distributed massive MIMO systems. The users transmit pilot-hopping sequences and the base stations use the received energy to detect the active users.  
We introduced user activity correlation where we modeled users that monitor physical phenomena and hence will activate based on their occurrence. 
Our proposed method is to add a regularization term to the NNLS. We studied two different regularizers inspired by total variance denoising and group LASSO. By simulating a realistic system, we showed that introducing regularizers that exploit the correlation greatly improves the performance of the user activity detection, both in terms of detecting the active users and estimating the position of the event that triggered the activation of the users. Which regularizer that is best will depend on the individual problem and model. 


\end{document}